\documentclass[10pt,letterpaper,twocolumn]{article} 

\usepackage{ol2}
\usepackage[draft]{hyperref}
\usepackage{amsmath}

\begin{document}

\twocolumn[ 

\title{Non-Hermitian laser arrays with tunable phase locking}


\author{Stefano Longhi}
\address{Dipartimento di Fisica, Politecnico di Milano and Istituto di Fotonica e Nanotecnologie del Consiglio Nazionale delle Ricerche, Piazza L. da Vinci 32, I-20133 Milano, Italy (stefano.longhi@polimi.it)}
\address{IFISC (UIB-CSIC), Instituto de Fisica Interdisciplinar y Sistemas Complejos, E-07122 Palma de Mallorca, Spain}

\begin{abstract}
Inspired by the idea of non-Hermitian spectral engineering and non-Hermitian skin effect, a novel design for stable emission of coupled laser arrays with tunable phase locking and strong supermode competition suppression is suggested. We consider a linear array of coupled resonators with asymmetric mode coupling displaying the non-Hermitian skin effect and show that, under suitable tailoring of complex frequencies of the two edge resonators, the laser array can stably emit in a single extended supermode with tunable phase locking and with strong suppression of all other skin supermodes. The proposed laser array design offers strong robustness against both structural imperfections of the system and dynamical instabilities typical of semiconductor laser arrays. 
 
\end{abstract}

 ] 

{\em Introduction.}
Recent technological advances in integrated active photonics and the introduction of new ideas from topological and non-Hermitian physics \cite{r1,r2,r3,r4,r5,r6,r7} have stimulated a renewed interest into the rather old problem of robust phase locking  of laser array systems \cite{r8,r9,r10,r11}. 
Coherent combination and phase locking of individual components are essential in obtaining high-radiance optical emitters and lasers \cite{r10,r11}, which are desirable for several applications ranging from
material processing, broad-area displays, industrial heating, and lidars. Integrated semiconductor laser arrays consist of a
large number of densely packed emitters which sustain several supermodes. It is well known that the nonlinear competition among the array
supermodes results in a complex spatiotemporal multimode 
oscillation \cite{r9,r12,r13} and corresponding 
incoherent emission, spectral broadening and poor beam quality. Several techniques of supermode selection and stabilization have been suggested and demonstrated in early works on laser arrays \cite{r14,r15,r16}.
Recently, new methods have been introduced to force laser emission in a single supermode inspired by ideas taken from other branches of physics, such as topological and non-Hermitian physics or supersymmetric quantum mechanics. Such methods include supersymmetric laser array design \cite{r17,r18,r19,r20}, topological  resonator cavities \cite{r21,r22,r23,r24,r25,r26,r27,r28,r29,r30,r31}, and non-Hermitian mode coupling \cite{r32,r33,r34,r35,r36}. In particular, arrays with asymmetric mode coupling realized via synthetic imaginary gauge fields have been predicted to display robustness against both structural imperfections (owing to the flowing nature of the supermode like in a topological laser \cite{r37}) and dynamical instabilities arising from the slow relaxation dynamics of the semiconductor gain medium \cite{r26,r32}. \\
In this Letter we suggest a simple design for robust phase locking of linear laser arrays with tunable phase locking condition, inspired by the concepts of non-Hermitian engineering and non-Hermitian skin effect \cite{r38,r39,r40,r41}. In a non-Hermitian lattice displaying the non-Hermitian skin effect \cite{r39,r40,r41} the energy spectrum and corresponding supermodes of the underlying Hamiltonian are strongly sensitive to the boundary conditions of the systems: while under periodic (ring) boundary conditions (PBC) the energy spectrum describes a closed loop in complex energy plane and the supermodes are extended waves, under open boundary conditions (OBC) the energy spectrum collapses to one or more open arcs in the interior of the PBC loop and the corresponding supermodes are squeezed toward the edges of the lattice (skin supermodes). The main idea is to harness such a strong boundary dependence of the energy spectrum and to engineer a linear chain of coupled resonators  displaying  a single low-loss extended supermode, with tailored phase locking condition picked up from one point of the PBC energy spectrum, and with all other OBC skin supermodes showing a higher loss rate. 

 {\it Non-Hermitian laser array engineering}. We consider an array of $N$ passive optical resonators with the same resonance frequency $\omega_0$ and loss rate $\gamma$ (i.e. complex frequency $\omega_0-i \gamma$), in which nearest-neighbor cavities are non-Hermitian coupled with asymmetric left/right coupling constants $\kappa_1=\kappa\exp(h)$ and $\kappa_2=\kappa \exp(-h)$, where $h$ is a synthetic imaginary gauge field.  Such a lattice  basically describes the clean (i.e. disorder-free) Hatano-Nelson  model \cite{r42} with a non-Hermitian Hamiltonian $\mathcal{H}$ {of elements $\mathcal{H}_{n,m}=\delta_{n,m-1} \kappa_1+ \delta_{n,m+1} \kappa_2$}. {We note that, contrary to other non-Hermitian lattice models exhibiting particle-hole symmetry and thus topologically-protected zero-energy modes \cite{r24,Referee}, the Hatano-Nelson model does not possess particle-hole-symmetry, rather it displays the non-Hermitian skin effect.}The experimental realization of such non-Hermitian asymmetric coupling has been reported in some recent works using integrated semiconductor microring laser technology \cite{r36,r37}. 
 For a purely passive system, the imaginary part of any eigenvalue of $\mathcal{H}$ must be negative (or vanishing) under different boundary conditions, which requires the minimal constraint $\gamma \geq 2 \kappa \sinh h$. For $h \neq 0$, the Hamiltonian displays the non-Hermitian skin effect, i.e. a strong dependence of its energy spectrum and corresponding eigenvectors on the boundary conditions \cite{r39,r40,r41}. We indicate by $\mathcal{H}_{PBC}$ and $\mathcal{H}_{OBC}$ the Hamiltonians under periodic (PBC) and open (OBC) boundary conditions, respectively, corresponding to a ring geometry and to a linear array geometry of the resonators [Figs.1(a) and (b)]. In the former case, the energy spectrum of $\mathcal{H}_{PBC}$ describes an ellipse in complex energy plane, $E_{PBC}(q)=-i \gamma+2 \kappa \cos(q-ih)$, and the corresponding eigenvectors (supermodes) are extended Bloch waves with quantized Bloch wave number $q=q_l= 2 \pi l/N$ ($l=0,1,2,...,N-1$). 
 {The various values of $q_l$ correspond to different phase locking conditions of the emitters with the same (homogeneous) intensity distribution in the near-field. The specific  value of $q_l$ determines the far-field pattern of the radiated light \cite{r8,r34}, with typically  a main single lobe at $q_l=0$ with the highest radiation efficiency and the formation of two lobes as $q_l$ is increased to $q_l= \pi$ (see Supplemental document).} In the open linear chain of Fig.1(b), the energy spectrum  collapses to a set of points on a segment in the interior of the ellipse, namely $E_{OBC}(q)=-i \gamma+2 \kappa \cos(q)$ with quantized values $q=q_l= l \pi/(N+1)$ ($l=1,2,...,N$); the corresponding supermodes are exponentially localized at the edge of the array (skin supermodes).  The localization properties of a given supermode with field amplitudes $\mathcal{E}_n$ is catched by the inverse participation ratio ${\rm IPR}=\sum_n (|\mathcal{E}_n|^4) / \left( \sum_n |\mathcal{E}_n|^2 \right)^2$, which varies  in the range $(0,1)$: an IPR value close to 1 corresponds to a tightly localized supermode, while a value of IPR close to zero corresponds to a fully extended supermode (in the large $N$ limit).
  When a uniform linear gain $g$ is added to each resonator by optical or electrical pumping, the linear coupled-mode equations describing the dynamics of the modal field amplitudes $\mathcal{E}_n$ in each resonator read
 \begin{equation}
 i \frac{d \mathcal{E}_n}{dt}= \sum_{l=1}^{N} \mathcal{H}_{n,l} \mathcal{E}_l+i g \mathcal{E}_n
 \end{equation}
 with $\mathcal{H}$ specified by the corresponding boundary conditions. Clearly, in the linear chain geometry of Fig.1(b), i.e. under OBC, all supermodes are degenerate in threshold, so that the array will exhibit rather generally a complex spatio-temporal multimode dynamics above threshold. Conversely, in the ring geometry of Fig.1(a) there is one supermode with lowest decay rate corresponding to the Bloch wave number (or phase locking condition) $q=\pi/2$: such a supermode will lase first near threshold, and under certain conditions it can be a stable attractor of the dynamics above threshold when the gain dynamics is considered \cite{r32}. The asymmetric coupling between adjacent resonators provides a chiral energy flow along the ring, making the phase locking robust against small-to-moderate structural imperfections or disorder in the system \cite{r37}. In the framework of the Hatano-Nelson model \cite{r42}, such a phase locking robustness is basically related to the phenomenon of non-Hermitian transparency \cite{r43}, i.e. the ability of light to flow immune through scattering centers in the lattice.  However, the supermode discrimination in the ring geometry becomes poor as the number $N$ of cavities is increased, resulting in a narrow stability region, and it is not possible to tune the phase locking condition far from $q=\pi/2$; a smaller value of $q$ would be more desirable to enhance the radiation efficiency in the far field \cite{r8}. \\
 \par
 In order to improve supermode discrimination and to tune the phase locking condition far from $q= \pi /2$, our main idea is to strategically design a third Hamiltonian $\mathcal{H}^{\prime}_{OBC}$, corresponding to a linear array with OBC but with modified complex frequencies $\delta \omega_1$ and $\delta \omega_N$ for the edge resonators of the array, as schematically shown in Fig.3(c). {i.e. 
 $(\mathcal{H}^{\prime}_{OBC})_{n,m}=(\mathcal{H}_{OBC})_{n,m}+\delta \omega_1 \delta_{n,1} \delta_{m,1}+ \delta \omega_N  \delta_{n,N} \delta_{m,N}$} (see also Supplemental document). In a practical setting, the control of the complex frequencies of the two edge resonators can be achieved by a judicious tuning of the optical (or current) pump rate and by a thermal tuning of the cavity resonances. Remarkably, after tuning $\delta \omega_1$ and $\delta \omega_N$ to 
 \begin{equation}
 {
\delta \omega_1=\kappa \exp(-h-iQ_0) \;, \;\; \delta \omega_n=\kappa \exp(h+iQ_0)}
\end{equation}
it can be proven (see Supplemental document) that the $N$ eigenvalues of $\mathcal{H}^{\prime}_{OBC}$ consist of a set of $(N-1)$ points on the straight segment with the same decay rate $\gamma$ as for $\mathcal{H}_{OBC}$, namely $E_l=-i \gamma+2 \kappa \cos ( \pi l /N)$ (for $l=1,2,3,...,N-1$), plus the additional isolated eigenvalue $E_0=-i \gamma+2 \kappa \cos (Q_0-ih)$, with a decay rate $\gamma_s=\gamma-2 \kappa \sin Q_0 \sinh h$. The corresponding supermodes are $(N-1)$ skin modes, squeezed toward the edge of the array, plus an extended (Bloch) supermode $\mathcal{E}_n=\exp(iQ_0 n)$ with phase locking condition defined by the angle $Q_0$. An example of energy spectrum and IPR of various supermodes of $\mathcal{H}^{\prime}_{OBC}$  is shown in Fig.1(c).
Clearly, for $h>0$ and $0<Q_0< \pi$ the extended Bloch supermode, whit the phase locking condition defined by the angle $Q_0$, shows the lowest oscillation threshold and it is thus expected to be the stable lasing attractor of the dynamics above threshold. Since the loss discrimination $\gamma-\gamma_s=2 \kappa \sin Q_0 \sinh (h)$ over all other supermodes is independent of array size $N$, the filtering method  holds for arbitrary long arrays. {Also, the method could be readily extended to a two-dimensional laser array (see Supplemental document)}.\\
\par

\begin{figure}[htb]
\centerline{\includegraphics[width=8.4cm]{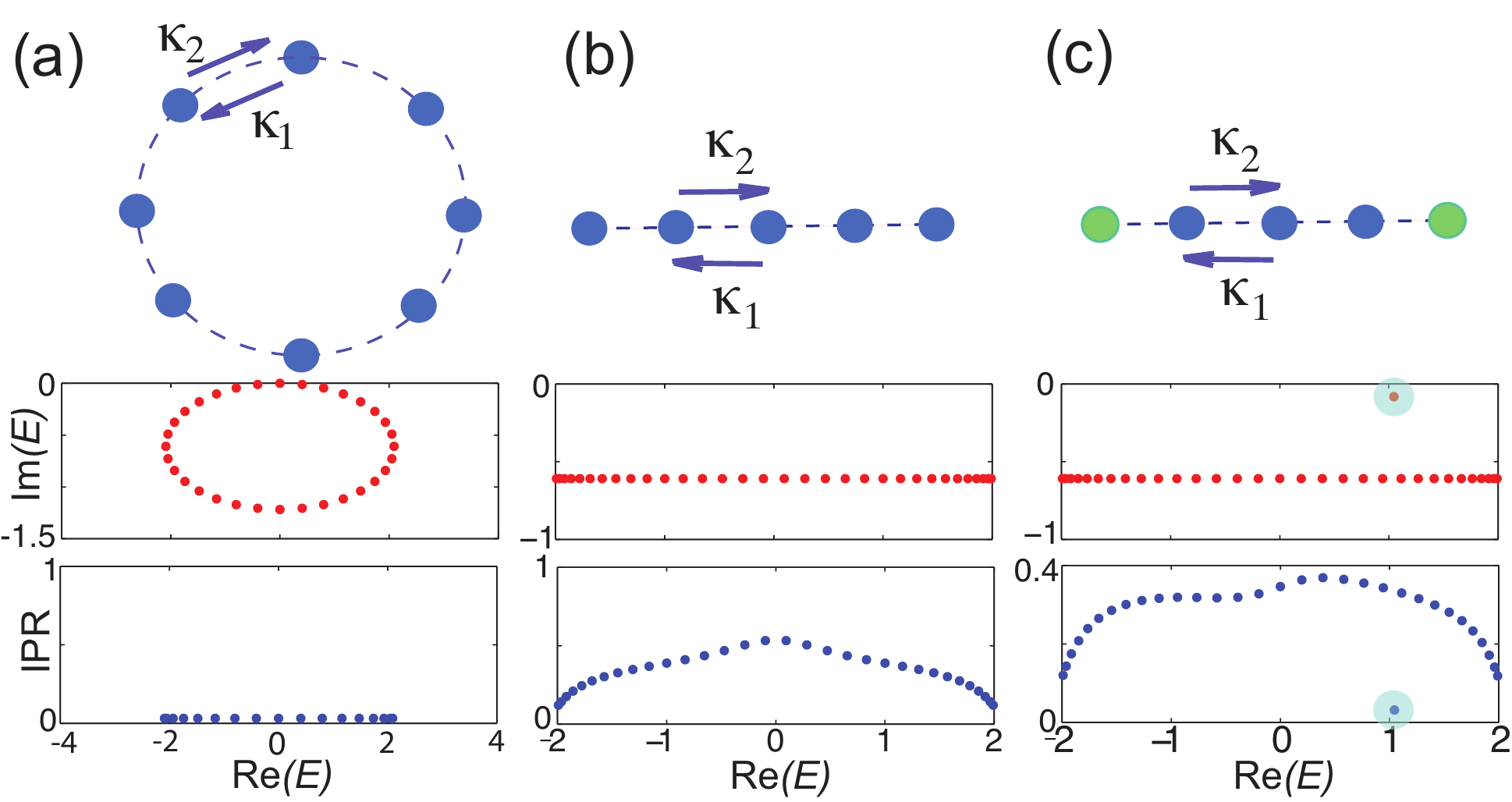}} \caption{
(Color online) Energy spectrum and IPR of corresponding supermodes in an array of coupled cavities with asymmetric coupling constants $\kappa_1=\kappa \exp(h)$, $\kappa_2=\kappa \exp(-h)$ under different geometric settings: (a) PBC (ring geometry), (b) OBC (linear chain), and (c) OBC with modified complex frequencies of edge resonators. The Hamiltonians of the three lattices are $\mathcal{H}_{PBC}$,  $\mathcal{H}_{OBC}$ and 
$\mathcal{H}_{OBC}^{\prime}$, respectively. Parameter values are: $h=0.3$, $N=32$, $\gamma / \kappa=2 \sinh (h) \simeq 0.609$, and $Q_0=\pi/3$ in (c). Energies are in units of $\kappa$. Note in (c) the existence of an isolated eigenenergy (highlighted by the shaded circle) corresponding to an extended supermode with the lowest IPR and decay rate.}
\end{figure} 
  \begin{figure}[htb]
 \centerline{\includegraphics[width=8.6cm]{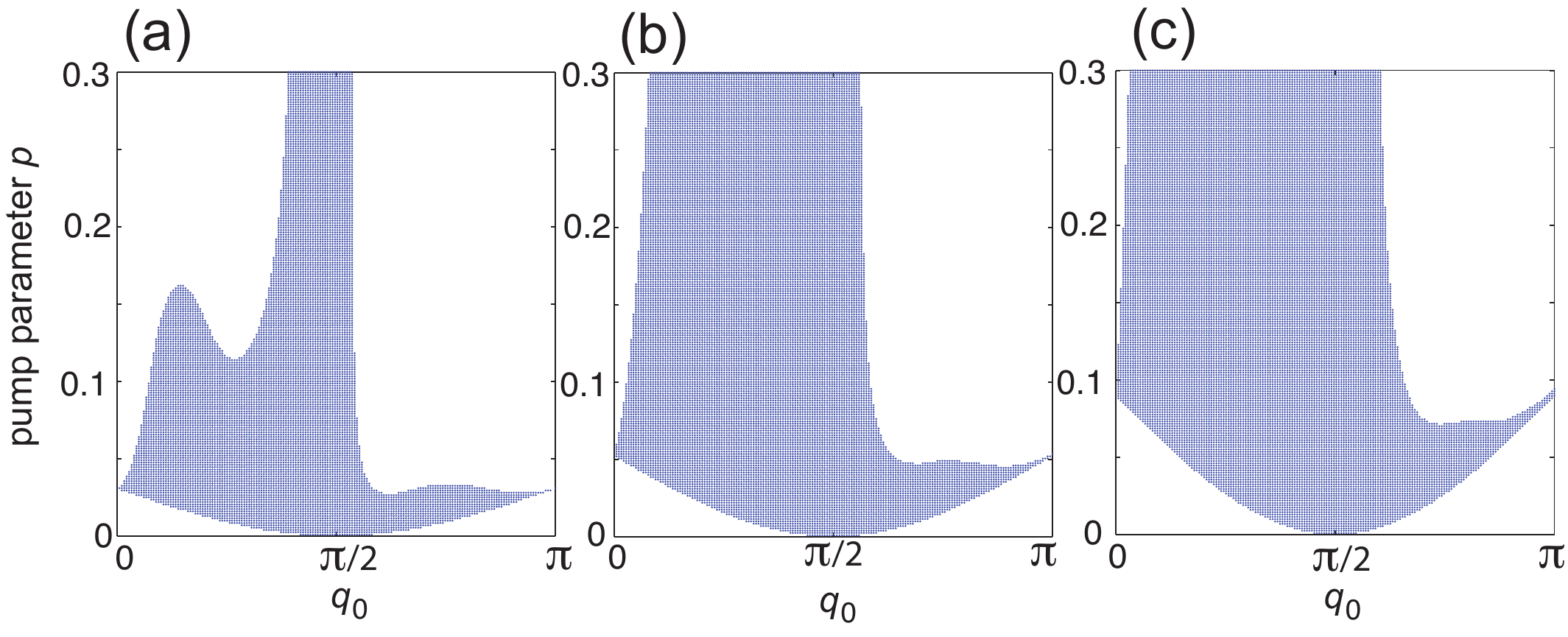}} \caption{
(Color online) Stability diagram (shaded area) in the $(Q_0,p)$ plane of the extended (Bloch) supermode of the laser array of Fig.1(c) comprising $N=32$ resonators  for a few increasing values of the imaginary gauge field $h$: (a) $h=0.3$, (b) $h=0.5$, and (c) $h=0.8$. Other parameter values are $ \tau_s / \tau_p= 2 \times 10^3$, $\kappa \tau_p=0.05$, $\alpha=3$, and $\gamma=2 \kappa \sinh h$. The lower boundary of the stability domain corresponds to the threshold value curve $p_{th}(Q_0)=\gamma_s \tau_p=2 \kappa \tau_p (1- \sin Q_0) \sinh h$.}
 \end{figure}  

 {\it Semiconductor laser dynamics.}  When the optical gain $g$ in the cavities of the array is provided by an inverted semiconductor medium, the spatio-temporal laser dynamics above threshold is described by a set of rate equations that account for nonlinear coupled dynamics of electric modal fields and carrier densities in the active (pumped) cavities. Such a rate equation model is capable of capturing the onset of dynamical instabilities arising from
 supermode competition and from nonlinear-induced resonance detuning of coupled cavities that can disrupt phase locking \cite{r9,r12,r13,r26,r32}. The laser rate equations read \cite{r13,r32,r33}
 \begin{eqnarray}
 \tau_p \frac{d\mathcal{E}_n}{dt} & = & (1-i \alpha) \mathcal{E}_n Z_n-i \tau_p \sum_{l=1}^N \mathcal{H}_{n,l}\mathcal{E}_l \\
 \tau_s \frac{dZ_n}{dt} & = & p-Z_n-(1+2 Z_n) |\mathcal{E}_n|^2
  \end{eqnarray}
  ($n=1,2,...,N$), 
where $\mathcal{E}_n$ is the normalized electric field amplitude in the $n$-th resonator of the array, $Z_n$ is the normalized excess carrier density, $\tau_p$ is the photon lifetime in each cavity due to strong material absorption when the semiconductor is not pumped, $\tau_s$ is the spontaneous carrier lifetime, $\alpha$ is the linewidth enhancement
factor, and $p$ is the normalized excess pump current, which provides a linear gain $g=p/ \tau_p$. In the engineered laser array of Fig.1(c), Eqs.(3) and (4) admit of the following above-threshold steady-steady solution, corresponding laser oscillation in the extended (Bloch) supermode
\begin{equation}
\overline{\mathcal{E}}_n=\mathcal{E}_0 \exp(i Q_0 n-i \Omega t) \; ,\;\; \overline{Z}_n= \tau_p \gamma_s \equiv {Z}_0
\end{equation}
where $\mathcal{E}_0=\{ (p-Z_0)/(1+2Z_0) \}^{1/2}$ is the homogeneous field amplitude and $\Omega=\alpha \gamma_s+2 \kappa \cosh (h) \cos (Q_0)$ the oscillation frequency offset from the passive resonator frequency $\omega_0$. The threshold pump rate is given by $p_{th}=Z_0= \tau_p \gamma_s$. 
The stability of the stationary supermode can be investigated by standard linear stability analysis \cite{r9,r13,r32}, which is detailed in the Supplemental document. In particular, the growth rate of perturbations can be numerically determined from the eigenvalues of a $3N \times 3N$ matrix. Examples of stability diagrams in the $(p,Q_0)$ plane, for a few increasing values of the imaginary gauge field $h$ and for values of $\tau_s / \tau_p$, $\alpha$ and $\kappa \tau_p$ typical of semiconductor laser array systems \cite{r32,r33}, are shown in Fig.2. 
\begin{figure}[htb]
\centerline{\includegraphics[width=8.6cm]{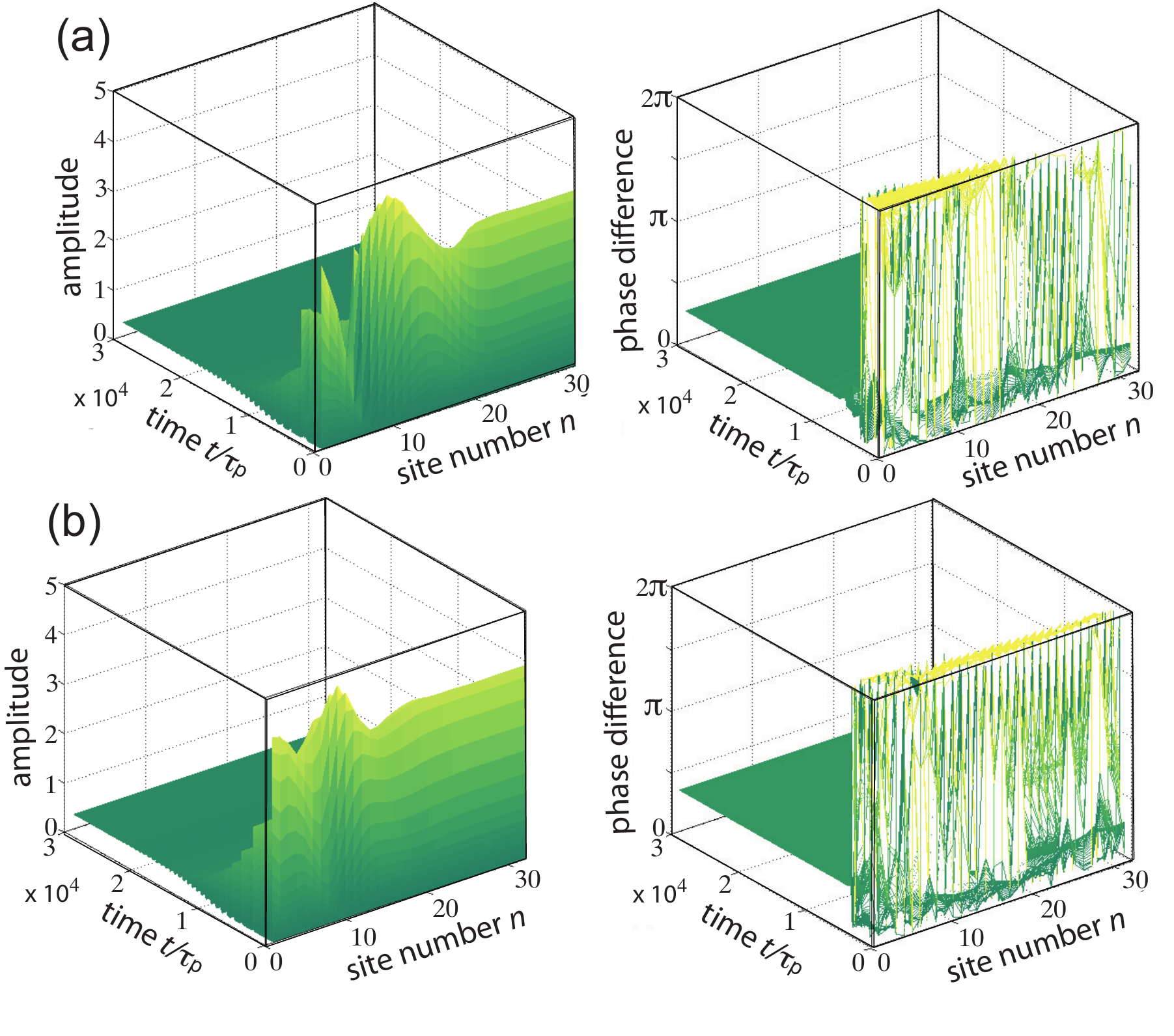}} \caption{ 
(Color online) Numerical simulations showing laser build up dynamics into the extended phase-locked Bloch supermode, starting from random small amplitudes of modes $\mathcal{E}_n$ and equilibrium carrier densities $Z_n=p$. The phase locking parameter is $Q_0= \pi/4$ in (a), and $Q_0= \pi/3$ in (b). Parameter values are $h=0.5$, $p=0.1$, $N=32$, $ \tau_s / \tau_p= 2 \times 10^3$, $\kappa \tau_p=0.05$, $\alpha=3$, and $\gamma / \kappa =2 \sinh h \simeq 1.042$.
The panels show the temporal behavior of amplitudes $|\mathcal{E}_n|$ and phase differences $\theta_n=(\varphi_{n+1}-\varphi_n)$ between adjacent resonators of the array for the normalized complex field amplitudes $\mathcal{E}_n=|\mathcal{E}_n| \exp(i \varphi_n)$.}
\end{figure} 
\begin{figure}[htb]
\centerline{\includegraphics[width=8.6cm]{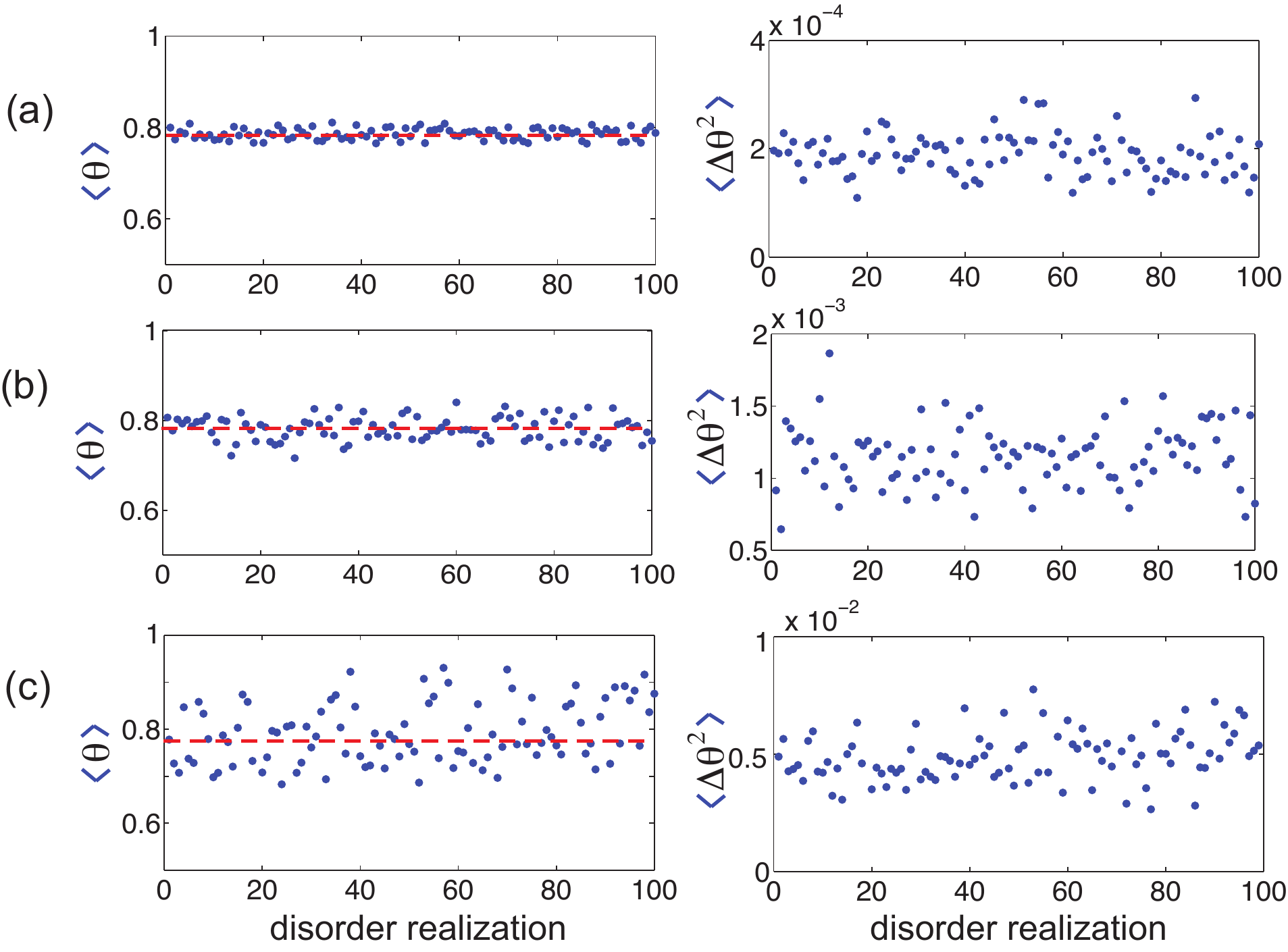}} 
\caption{ 
(Color online)  Robustness of phase locking of the extended Bloch supermode in the presence of disorder in the cavity resonance frequencies. The plots show the behavior of the mean phase difference $\langle \theta \rangle$ (left panels) and variance $\langle \Delta \theta^2 \rangle$ (right panels) for 100 different realizations of disorder of the cavity resonance frequencies $\delta \Omega_n= \kappa \sigma_n$ for a few increasing values of disorder strength $\sigma$: (a) $\sigma=0.2$, (b) $ \sigma=0.5$, and (c) $\sigma=1$. Parameter values are as in Fig.3(a). The dashed horizontal line in the left panels corresponds to the ideal phase locking condition $Q_0 =\pi /4$ in the disorder-free array.}
\end{figure} 
Note that the stability domain widens around $Q_0= \pi/2$ as $h$ is increased, corresponding to an enhanced suppression of skin supermodes. Note also the strong asymmetric behavior of the stability for $Q_0> \pi/2$ and $Q_0< \pi/2$, with a narrower stability region in the $Q_0> \pi/2$ region. This behavior is ascribable to the gain-induce frequency shift via the linewidth enhancement factor $\alpha>0$, which tends to rapidly destabilize the supermodes with $Q_0> \pi/2$ as compared to the ones with $Q_0 < \pi/2$. A reversed behavior, i.e. a narrower stability region for $Q_0< \pi/2$, would be observed by flipping the sign of $\alpha$. 
 Figures 3(a) and 3(b) show typical examples of laser built up from initial noise and stable oscillation in the dominant supermode, after relaxation oscillation transient, for two different values of $Q_0$, obtained by proper tailoring the complex frequencies of edge resonators according to Eq.(2). The results are obtained by direct numerical simulations of Eqs.(3) and (4) using an accurate fourth-order Runge-Kutta method;  initial condition is a small random noise of field amplitudes $\mathcal{E}_n$ and equilibrium carrier densities. \\
 The lasing supermode turns out to be stable against small-to-moderate disorder in the system, a robustness which is related to the phenomenon of non-Hermitian transparency discussed in Refs.\cite{r37,r43}. As an example, Fig.4 illustrates the robustness of the phase locking condition in the presence of disorder of the resonance frequencies of the resonators, which are varied from the reference and common value $\omega_0$ by a quantity $\delta \Omega_n= \sigma_n \kappa$, where $\sigma_n$ are independent random variables with the same probability density distribution. Specifically, we assume for $\sigma_n$ a uniform distribution in the range $(-\sigma /2, \sigma /2)$, where the dimensionless parameter $\sigma$ measures the strength of disorder with respect to the coupling constant $\kappa$. The figures depict the behavior of the mean $\langle \theta \rangle = (\sum_n \theta_n)/(N-1)$ and variance $\langle \Delta \theta^2 \rangle=  (\sum_n ( \theta_n-\langle \theta \rangle )^2)/(N-1)$ of the phase differences $\theta_n = (\varphi_{n+1}-\varphi_n)$ between adjacent resonators in the array, after the initial transient laser switch on dynamics for the same parameter values as Fig.3(a) and for 100 different realizations of disorder. Note that, even for a moderate disorder strength of cavity resonances, comparable to the coupling constant $\kappa$ [Fig.4(c)], the phase locking regime is not destabilized. A similar behavior is found when considering disorder in the coupling constants of the array, {or simultaneous disorder in both real and imaginary parts of the resonance frequencies (see Supplemental document)}. \\
 \\
 {\it Conclusion.} A method for robust and tunable phase locking of laser arrays, inspired by the concepts of non-Hermitian engineering and non-Hermitian skin effect, has been theoretically suggested. As compared to other methods based  on the use of topological or supersymmetric cavities, our technique enables for tuning of the laser phase locking condition, avoiding the onset of dynamical instabilities typical of semiconductor laser arrays. The present results provide important insights into the design of laser arrays, suggesting a potentially powerful application of the recently-introduced concept of non-Hermitian skin effect \cite{r39,r40,r41}, and are expected to stimulate further experimental and theoretical studies in the rapidly growing research areas of active integrated photonics and non-Hermitian optics. \\


\begin{thebibliography}{99}



\bibitem{r1}
L. Lu, J.D. Joannopoulos, and M. Solja\v{c}ic, Nat. Photon. {\bf 8}, 821 (2014).
\bibitem{r2}
A.B. Khanikaev and G. Shvets, Nat. Photon. {\bf 11}, 763 (2017).
\bibitem{r3}
 T. Ozawa, H. M. Price, A. Amo, N. Goldman, M. Hafezi, L. Lu, M. C. Rechtsman, D. Schuster, J. Simon, O. Zilberberg, and I. Carusotto, Rev. Mod. Phys. {\bf 91}, 015006 (2019).
 \bibitem{r4}
 L. Feng, R. El-Ganainy, and  L. Ge,  Nature Photon. {\bf 11}, 752 (2017).
 \bibitem{r5}
R. El-Ganainy, K.G. Makris, M. Khajavikhan, Z.H. Musslimani, S. Rotter, and D.N. Christodoulides, Nature Phys. {\bf 14}, 11 (2018).
\bibitem{r6}
B. Midya, H. Zhao, and L. Feng, Nature Commun. {\bf 9}, 2674 (2018).
\bibitem{r7}
S. Longhi, EPL {\bf 120}, 64001 (2017).
\bibitem{r8}
E. Kapon, J. Katz, and A. Yariv, Opt. Lett. {\bf 10}, 125 (1984).
\bibitem{r9}
 H.G. Winful and S.S. Wang, Appl. Phys. Lett. {\bf 53}, 1894 (1988).
 \bibitem{r10}
A. F. Glova, Quantum Electron. {\bf 33}, 283 (2003).
\bibitem{r11}
T.Y. Fan, IEEE J. Sel. Top. Quantum Electron. {\bf 11}, 567 (2005).
\bibitem{r12}
 H.G. Winful and L. Rahman, Phys. Rev. Lett. {\bf 65}, 1575 1990).
 \bibitem{r13}
 R.d. Li and T. Erneux,  Phys. Rev. A {\bf 46}, 4252 (1992).
 \bibitem{r14}
C. P. Lindsey, E. Kapon, J. Katz, S. Margalit, and A. Yariv,  Appl. Phys. Lett. {\bf 45}, 722 (1984).
\bibitem{r15}
J. Katz, S. Margalit, and A. Yariv,  Appl. Phys. Lett. {\bf 42}, 554 (1983).
\bibitem{r16}
J. R. Leger, M. L. Scott and W. B. Veldkamp,  Appl. Phys. Lett. {\bf 52}, 1771 (1988).
\bibitem{r17}
M.P. Hokmabadi, N.S. Nye, R. El-Ganainy, D.N. Christodoulides, and M. Khajavikhan, Science {\bf 363}, 6427 (2019).
\bibitem{r18}
B. Midya, H. Zhao, X. Qiao, P. Miao, W. Walasik, Z. Zhang, N.M. Litchinitser, and L. Feng,  Photon. Res. {\bf 7}, 363 (2019).
\bibitem{r19}
D.A. Smirnova, P. Padmanabhan, and D. Leykam, Opt. Lett. {\bf 44}, 1120 (2019).
\bibitem{r20}
X. Qiao, B. Midya, Z. Gao, Z. Zhang, H. Zhao, T. Wu, J. Yim, R. Agarwal, N.M. Litchinitser, and L. Feng, Science {\bf 372}, 403 (2021).
 \bibitem{r21}
 B. Bahari, A. Ndao, F. Vallini, A. El Amili, Y. Fainman, and B. Kante, Science {\bf 358},  636 (2017).
 \bibitem{r22}
 P. St-Jean, V. Goblot, E. Galopin, A. Lemaitre, T. Ozawa, L. Le Gratiet, I. Sagnes, J. Bloch, and A. Amo,
Nature Photon. {\bf 11}, 651 (2017).
 \bibitem{r23}
H. Zhao, P. Miao, M.H. Teimourpour, S. Malzard, R. El-Ganainy, H. Schomerus, and L. Feng, Nature Commun. {\bf 9}, 981 (2018).
 \bibitem{r24}
M. Pan, H. Zhao, P. Miao, S. Longhi, and L. Feng, Nature Commun. 9, 1308 (2018).
\bibitem{r25}
M.A. Bandres, S. Wittek, G. Harari, M. Parto, J. Ren, M. Segev, D.N. Christodoulides, and M. Khajavikhan, Science {\bf 359}, eaar4005 (2018).
\bibitem{r26}
S. Longhi, Y. Kominis, and V. Kovanis, EPL {\bf 122}, 14004 (2018).
\bibitem{r27}
S. K. Ivanov, Y. Q. Zhang, Y. V. Kartashov, and D. V. Skryabin, APL Photonics {\bf 4}, 126101 (2019).
\bibitem{r28}
D. Smirnova, A. Tripathi, S. Kruk, M.-S. Hwang, H.-R. Kim, H.-G. Park, and Y. Kivshar, Light: Sci. Appl. {\bf 9}, 127 (2020).
\bibitem{r29}
Y. Q. Zeng, U. Chattopadhyay, B. F. Zhu, B. Qiang, J. H. Li, Y. H. Jin, L. H. Li, A. G. Davies, E. H. Linfield, B. L. Zhang, Y. D. Chong, and Q. J. Wang, Nature {\bf 578}, 246 (2020).
\bibitem{r30}
S. Gundogdu, J. Thingna, and D. Leykam, Opt. Lett. {\bf 45}, 3673 (2020).
\bibitem{r31}
A. Dikopoltsev, T. H. Harder, E. Lustig, O. A. Egorov, J. Beierlein, A. Wolf, Y. Lumer, M. Emmerling, C. Schneider, S. Hofling, M. Segev, and S. Klembt, Science {\bf 373}, 1514 (2021).
\bibitem{r32}
 S. Longhi and L. Feng, APL Photonics {\bf 3}, 060802 (2018).
 \bibitem{r33}
S. Longhi, Ann. Phys. {\bf 530}, 1800023 (2018).
\bibitem{r34}
J. Ding and M.-A. Miri, Opt. Lett. {\bf 44}, 5021 (2019).
\bibitem{r35}
Z. Zhang, X. Qiao, B. Midya, K. Liu, J. Sun, T. Wu, W. Liu, R. Agarwal, J.M. Jornet, S. Longhi, N.M. Litchinitser, and L. Feng, Science {\bf 368}, 760 (2020).
\bibitem{r36}
Y.G.N. Liu, O. Hemmatyar, A.U. Hassan, P.S. Jung, J.-H. Choi, D.N. Christodoulides, and   M. Khajavikhan, APL Photonics {\bf 6}, 050804 (2021).
\bibitem{r37}
S. Longhi, D. Gatti, and G. Della Valle, Sci. Rep. {\bf 5}, 13376 (2015).
\bibitem{r38}
D. Leykam, K. Y. Bliokh, C. Huang, Y. D. Chong, and F.
Nori, Phys. Rev. Lett. {\bf 118}, 040401 (2017).
\bibitem{r39}
S. Yao and Z. Wang, Phys. Rev. Lett. {\bf 121}, 086803 (2018).
\bibitem{r40}
E.J. Bergholtz, J.C. Budich, and F.K. Kunst,  Rev. Mod. Phys. {\bf 93}, 015005 (2021).
\bibitem{r41}
N. Okuma, K. Kawabata, K. Shiozaki, and M. Sato, Phys. Rev. Lett. {\bf 124}, 086801 (2020).
\bibitem{r42}
N. Hatano and D.R. Nelson, Phys. Rev. Lett. {\bf 77}, 570 (1996).
\bibitem{Referee}
{
F. Hentinger, M. Hedir, B. Garbin, M. Marconi, L. Ge, F. Raineri, J.A. Levenson, and A.M. Yacomotti,
Phot. Res. {\bf 10}, 574 (2022).}
\bibitem{r43}
S. Longhi, D. Gatti and G. Della Valle, 
Phys. Rev. B {\bf 92}, 094204 (2015).


\end{thebibliography}
\end{document}